\documentclass[]{imag-ms-template}

\usepackage{siunitx}
\usepackage{lineno}
\usepackage{url}

\title{Functional Ultrasound Imaging Combined with Machine Learning for Whole-Brain Analysis of Drug-Induced Hemodynamic Changes}

\author{Jared Deighton$^{1,\dagger}$,
Shan Zhong$^{2,3,\dagger}$,
Kofi Agyeman$^{3,4,5}$,\\
Wooseong Choi$^{5}$,
Charles Liu$^{3,5,6,7}$,
Darrin Lee$^{3,5,6,7}$,\\
Vasileios Maroulas$^{1\ast}$,
Vassilios Christopoulos$^{2,3,4,5,6\ast}$ \\
\small{$^1$Department of Mathematics, University of Tennessee, Knoxville, Knoxville, TN, USA;}\\
\small{$^2$Neuroscience graduate program, University of California Riverside, Riverside, CA, USA;}\\
\small{$^3$Alfred E.Mann Department of Biomedical Engineering, University of Southern California, Los Angeles, CA, USA;}\\
\small{$^4$ Department of Bioengineering, University of California Riverside, Riverside, CA, USA;}\\
\small{$^5$ Dept. of Neurological Surgery, Keck School of Medicine, University of Southern California, Los Angeles, CA, USA;}\\
\small{$^6$ Neurorestoration Center, Keck School of Medicine, University of Southern California, Los Angeles, CA, USA;}\\
\small{$^7$ Rancho Los Amigos National Rehabilitation Center, Downey, CA, USA;}\\
\small{$^\dagger$These authors contributed equally to this work.}\\
\small{$\ast$These authors jointly supervised this work (Correspondence)}\\
\small{$\ast$ E-mail: Corresponding vchristo@usc.edu, vmaroula@utk.edu}}

\begin{document}

\maketitle


\begin{abstract}
Functional ultrasound imaging (fUSI) is a cutting-edge technology that measures changes in cerebral blood volume (CBV) by detecting backscattered echoes from red blood cells moving within its field of view (FOV). It offers high spatiotemporal resolution and sensitivity, allowing for detailed visualization of cerebral blood flow dynamics. While fUSI has been utilized in preclinical drug development studies to explore the mechanisms of action of various drugs targeting the central nervous system, many of these studies rely on predetermined regions of interest (ROIs). This focus may overlook relevant brain activity outside these specific areas, which could influence the results. To address this limitation, we compared three machine learning approaches—convolutional neural network (CNN), support vector machine (SVM), and vision transformer (ViT)—combined with fUSI to analyze the pharmacodynamics of Dizocilpine (MK-801), a potent non-competitive NMDA receptor antagonist commonly used in preclinical models for memory and learning impairments. While all three machine learning techniques could distinguish between drug and control conditions, CNN proved particularly effective due to their ability to capture hierarchical spatial features while maintaining anatomical specificity. Class activation mapping revealed brain regions, including the prefrontal cortex and hippocampus, that are significantly affected by drug administration, consistent with the literature reporting a high density of NMDA receptors in these areas. Overall, the combination of fUSI and CNN creates a novel analytical framework for examining pharmacological mechanisms, allowing for data-driven identification and regional mapping of drug effects while preserving anatomical context and physiological relevance.
\end{abstract}

\section{Introduction}
Functional ultrasound imaging (fUSI) is an emerging hemodynamic-based neuroimaging technology that measures cerebral blood volume (CBV) changes by detecting backscattered echoes from red blood cells moving within its field of view (FOV) (\cite{mace2011functional,mace2013functional}). It provides a unique combination of large spatial coverage, high spatiotemporal resolution ($\sim$ 100 \textmu m and $\sim$ 100 ms) and sufficient sensitivity to detect slow blood flow changes (less than 1 mm/s). The relative simplicity and portability of ultrasound scanners have allowed fUSI to be performed in a wide range of preclinical and clinical studies, providing minimally invasive neural imaging in species ranging from mice to humans (\cite{Osmanski2014odor, osmanski2014functional, Imbault2017intra, demene2017functional, norman2021single, griggs2023decoding}). Among its various applications, fUSI has been employed in preclinical drug development to elucidate the mechanisms of action of drugs targeting the central nervous system. In particular, the mechanisms of various drugs, including anesthetics, cholinesterase agonists and antagonists, N-methyl-D-aspartate (NMDA) receptor antagonists, and selective norepinephrine reuptake inhibitors (SNRIs), have been investigated to better understand their effects on brain function and vascular dynamics, and to help accelerate the development of new drug therapies (\cite{rabut2020pharmaco, Vidal2020pharmacofUS, Vidal2020drug,crown2024theta}).

Despite the important contribution of these neuropharmacological studies, they primarily examine the effects of drugs on hemodynamic signals within specific regions of interest (ROIs) and/or the functional connectivity between these regions. This approach may introduce bias by neglecting drug-induced changes in neural activity occurring outside these targeted areas. Therefore, there is a critical need to develop advanced analytical tools capable of assessing the dynamic effects of drugs on the brain without relying on predefined region specification. In this study, we develop and compare multiple machine learning approaches—convolutional neural network (CNN), support vector machine (SVM), and vision transformers (ViT)—combined with fUSI to elucidate the pharmacodynamics of dizocilpine (MK-801) in the brain. MK-801 is a potent and selective NMDA receptor antagonist originally used as a pharmacological model of psychosis in rodents (\cite{Andine1999}), and is still extensively employed to model schizophrenia (\cite{Zepeda2022}). Our comparative analysis showed that although CNN and ViT achieve similar classification performance—with SVM exhibiting comparatively lower accuracy—they differ notably in biological interpretability. When combined with axiom-based grad-class activation mapping (XGrad-CAM) (\cite{fu2020axiom}), CNN revealed anatomically specific activation patterns in cortical and hippocampal regions with high NMDA receptor expression (\cite{watanabe1994distinct}). Post-hoc quantitative analysis of CBV changes in these regions confirmed that MK-801 administration caused significant hemodynamic reduction, validating that CNN can accurately detect and localize drug effects while maintaining interpretability. Overall, the combination of fUSI technology with CNN provides a powerful new framework for investigating pharmacological mechanisms in the brain and has potential applications for accelerating drug development.

\section{Materials and methods}

\subsection{Animals}
Twenty-three (23) 8-12-week-old male mice were used in this study (C57BL/6, Charles River Laboratories, Hollister, CA). All mice were group-housed, fed \textit{ad libitum}, and maintained at a regular light-dark cycle of 12 hours. The animals were divided into two groups: MK-801 drug group (n=10) and saline control group (n=13).  

\subsection{Surgical Procedures}
The mice were anesthetized using 5\% isoflurane solution delivered in a mixture of oxygen and nitrous oxide (1:2 ratio) and then maintained at a constant isoflurane concentration of 1.5-2\%  throughout the experiment. Body temperature was kept stable using an electric warming pad. The animals were head-fixed in a stereotaxic frame with ear bars to minimize head movement and reduce motion artifacts. A commercially available depilatory cream (Nair, Pharmapacks) was utilized to remove hair from the scalp followed by application of an echographic ultrasonic gel on the intact scalp-skin to enhance acoustic coupling during fUSI signal acquisition. All the experimental and surgical protocols were approved by the Institutional Animal Care and Use Committee of the University of Southern California (IACUC \#21006).

\subsection{Data Acquisition}
\begin{figure}[h!t]
    \centering
    \includegraphics[width = \textwidth]{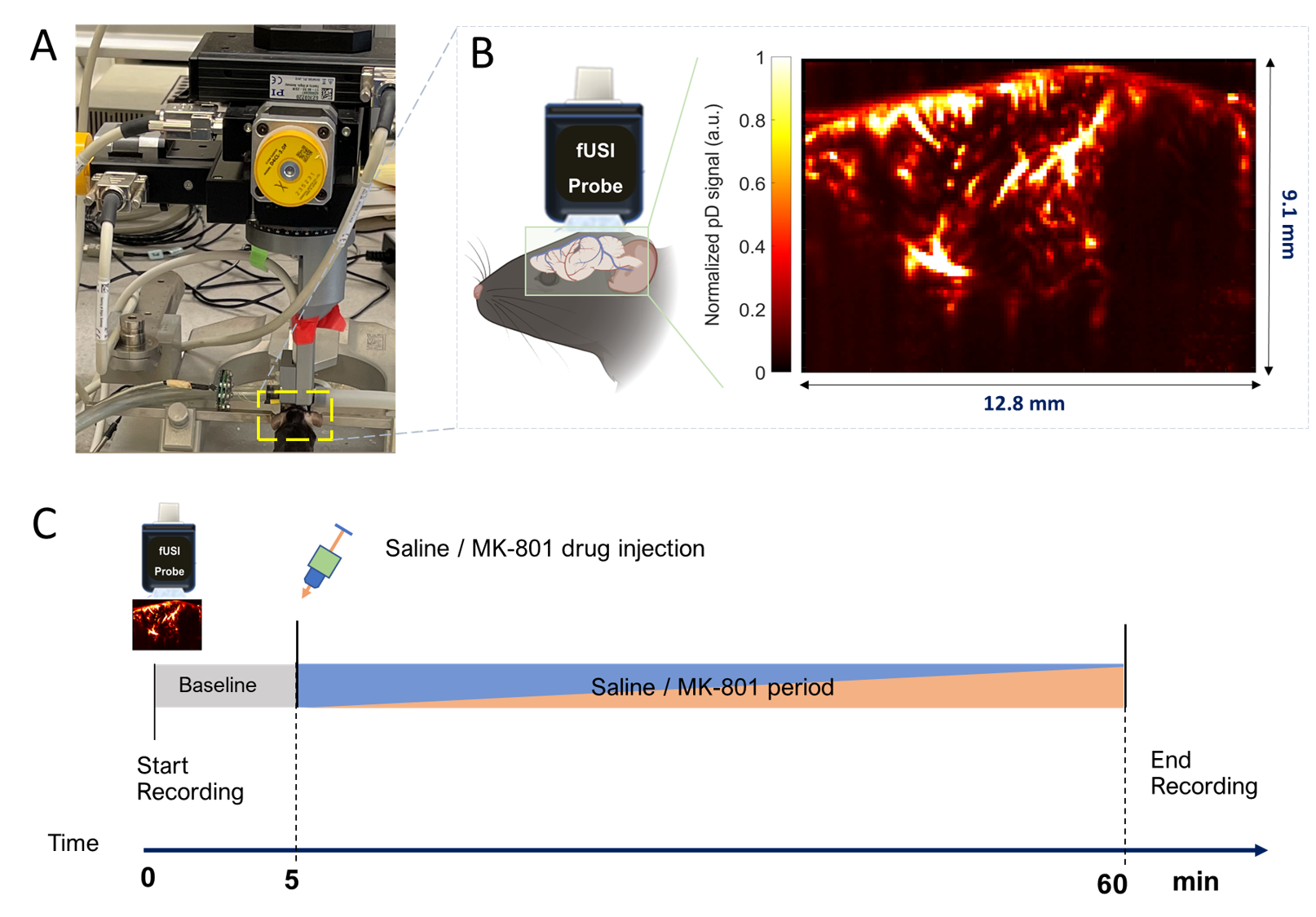}
    \caption{ Experimental setup and fUSI acquisition.
     (\textbf{A}) fUSI motorized system setup for signal acquisition. (\textbf{B}) Power Doppler-based 2D vascular map image of a mouse brain in a sagittal plane. (\textbf{C}) Diagram of the experimental protocol for the 60 min of continuous fUSI acquisition. After 5 min, saline or 1.5 mg/kg MK-801 was intraperitoneally injected to the animals. The blue/orange color means the mouse is either injected with MK-801 or saline.}
     \label{fig:ExpSetup}
\end{figure}

Transcranial power Doppler (pD) images were obtained using the Iconeus One scanner (Iconeus, Paris, France). A 128-element linear array transducer probe with a \SI{15.6}{\mega\hertz}  center frequency and 0.1 mm pitch was placed on intact mouse skulls. The probe was fixed in a motorized system during the course of the experiment (Fig.~\ref{fig:ExpSetup}A). Prior to recording, the target sagittal plane for imaging was determined by performing a 3-D whole-brain fUSI scan for each animal. The plane was then aligned with the standard Allen Mouse Common Coordinate Framework brain atlas using the specialized software provided with the Iconeus One system (\cite{wang2020allen}). The details of fUSI parameters are described in (\cite{crown2024theta}). Briefly, each image was constructed from 200 compounded frames, acquired at \SI{500}{\hertz} using 11 tilted plane waves (\SI{-10}{\degree} to \SI{+10}{\degree}, \SI{2}{\degree} increments). The pulse repetition frequency was \SI{5.5}{\kilo\hertz}, with continuous acquisition of \SI{400}{\milli\second} blocks of compounded images, separated by \SI{600}{\milli\second} intervals. This approach enables pD image acquisition with in-plane spatial resolution of 100 \textmu m × 100 \textmu m, slice thickness of 400 \textmu m, FOV of 12.8 mm in width and 10.0 mm in depth, and an overall image-frame production rate of \SI{1}{\hertz} (Fig.~\ref{fig:ExpSetup}B). The fUSI data acquisition protocol consisted of 5 minutes of pre-injection recording, followed by intraperitoneal (i.p.) injection of either 0.2 ml of saline or MK-801 (1.5 mg/kg) at the 5 minutes mark, followed by an additional 55 minutes of recording post-saline/MK-801 injection (Fig.~\ref{fig:ExpSetup}C). For data analysis purposes, only the final two minutes of the pre-injection period (minutes 3-5) were used as the baseline reference, as the signal had stabilized during this interval.

\subsection{Data Pre-processing and Registration}
We implemented the NoRMCorre motion correction technique (\cite{pnevmatikakis2017normcorre}), combined with our in-house algorithms for filtering breathing and high-frequency signal to reduce noise. Specifically, we applied a lowpass filter featuring a normalized passband frequency of \SI{0.02}{\hertz} and a stopband attenuation of 60 dB, which also offsets any delay caused by the filtering process, thereby eliminating high-frequency fluctuations.

Furthermore, all images were registered to a common reference image to prevent the machine learning models from being influenced by inter-subject variability in brain structures. We first identified the reference animal based on the imaging quality, selecting one with clearly defined subcortical structures. Then we computed the mean image of the first ten images at the beginning of the third minute of the 5-minute pre-injection baseline period to create a stable fixed image (reference image), in order to mitigate any potential variability that might arise from using a single image as a reference. We then employed the {\it Imregdeform} algorithm, which is part of the Matlab Image Processing Toolbox (since 2022b). This algorithm employs a parametric approach to non-rigid image registration with total variation regularization. All images were aligned to the reference frame using the {\it Imregdeform} algorithm, which aids in simplifying subsequent data analysis.The default parameters were used for {\it Imregdeform}.

\subsection{Models and Feature selection}

\subsubsection{Implementation and Parameter Selection}

SVM was implemented using scikit-learn (\cite{scikit-learn}), while CNNs and ViTs were implemented using PyTorch (\cite{paszke2019pytorch}). Most of the code was executed on a \SI{2.8}{\giga\hertz} Quad-Core Intel Core i7 processor with 16 GB of memory. Parameter sweeps were run using a Tesla K80 GPU with 64GB of memory. Parameter sweep was performed using Weights and Biases (\cite{wandb}) grid sweep. After the grid sweep, we subjected the 15 best hyperparameter combinations to 10-fold cross-validation, from which the best model  - i.e., the model yielding the highest peak accuracy - was chosen.

\subsubsection{Convolutional neural network (CNN)}
The CNN progressively reduces the spatial dimensions of the image input while increasing the depth of feature maps through multiple convolutional layers. To ensure compatibility with class activation mapping, the architecture includes a global average pooling (GAP) layer followed by a single linear layer, which outputs two features corresponding to the binary classification task.
We conducted extensive parameter sweeps to determine the optimal architecture and hyperparameters (Table \ref{table:CNN_hyperparameters}, best-performing values in bold). The final architecture consists of four convolutional layers with 32 filters each, employing the $ELU(\cdot)$  activation function. The $ELU(\cdot)$ activation function, which was selected over ReLU and LeakyReLU alternatives,  is defined by:
\begin{gather}
    ELU(x) = \begin{cases}
        x & \text{if}\ x > 0 \\
        \exp(x) - 1 & \text{if}\ x \leq 0 
    \end{cases}
\end{gather}

This activation function is known to speed up learning and lead to higher classification accuracies by pushing the network's mean activation toward zero (\cite{clevert2015fast}). The model is trained using the Adam optimizer with a learning rate of 1e-3 and batch size of 32.

\begin{table}[ht!]
\centering
\begin{threeparttable}

\caption{Hyperparameters used in the CNN parameter sweep.}
\label{table:CNN_hyperparameters}

\begin{tabular}{ll}
\toprule
\textbf{Hyperparameter}        & \textbf{Value}               \\ \hline
Batch size   & \textbf{32}, 64                            \\
Learning rate & 1e-4, \textbf{1e-3}, 1e-2                        \\
CNN nonlinearity  & ReLU$(\cdot)$, $\mathbf{ELU}(\cdot)$, LeakyReLU$(\cdot)$                        \\
Number of Filters & \textbf{32}, 64
\\
Number of Convolutional Layers & 2, 3, \textbf{4}\\
Optimizer & \textbf{Adam} (\cite{kingma2014adam}) 
\\
Visualization Technique & \textbf{XGRAD-CAM} (\cite{lu2017automatic}) \\
Loss Function & \textbf{Cross-entropy} \\ 
\bottomrule
\end{tabular}
\begin{tablenotes}
\small
\centering
\item \textbf{Bold values} indicate hyperparameters used in the final selected model.
\end{tablenotes}
\end{threeparttable}
\end{table}

\subsubsection{Class activation mapping} 

Class activation mapping was performed using XGRAD-CAM (\cite{lu2017automatic}) with a sliding time window of one minute. The localization map for a given input is computed as follows
\begin{gather}
\label{eq:locatization}
    L^{(c)}(x,y) = ReLU\left( \sum_k w_k^{(c)} A_k(x,y) \right)
\end{gather}
where $ReLU(x) = \max(0, x)$ and the coefficient $ w_k^{(c)}$ being defined as 
\begin{gather}
    \label{eq:weights}
     w_k^{(c)} = \sum_{i = 1}^{H} \sum_{j = 1}^W \left(\frac{\partial Y^{(c)}}{\partial A_k(i,j)} \cdot \frac{A_k(i,j)}{\sum_{m = 1}^{H} \sum_{n = 1}^W A_k(m,n)} \right)
\end{gather}
where $A_k(x,y)$ is the activation of node $k$ in the target layer of the model at position $(x,y)$ and $Y^{(c)}$ is the model output score for class $c$ before activation. Here, $H$ and $W$ are the height and width of the feature maps, respectively. 

The target layer for this analysis was chosen to be the final convolutional layer, which contains higher-level feature representations and is commonly chosen for CAM (\cite{zhou2016learning}). 

The weights $w_k^{(c)}$ found in Equation \eqref{eq:weights} were computed using the product of the gradient of the model's raw score for the class $(c)$ with respect to each activation and a normalization term. This encodes the importance of each activation in the final convolutional layer regarding the final classification decision for class $(c)$. In our case, the final convolutional layer produces 128 feature maps with height, $H = 16$, and width $W = 12$. 

As described in Equation \eqref{eq:locatization}, after calculating the weights for the given class and node, the weighted average was computed and the $ReLU(\cdot)$ activation function was applied to obtain the activation at each region. The result is a 12 $\times$ 16 activation map, corresponding to the dimensions of the feature maps from the final convolutional layer. For better comparison with the original image, the 12 $\times$ 16 image is up-sampled to the original image size of 91 $\times$ 128 via bilinear interpolation. 

\subsubsection{Vision transformer (ViT)}

ViTs are deep learning models designed for computer vision tasks (\cite{alexey2020image}). Unlike CNNs, ViTs process images by first splitting them into fixed-size patches and then applying a transformer architecture to the sequence of these patches. The architecture details of our ViT implementation are determined through parameter sweeps (Table \ref{table:VIT_hyperparameters}, best-performing values in bold). The final model uses 16 transformer layers with 4 attention heads and an embedding dimension of 256, trained with a batch size of 16 and learning rate of 1e-4.

To visualize the regions important for classification, we employed Attention Rollout (\cite{abnar2022quantifying}) with max fusion and a discard ratio of 0.9. For an in-depth description of attention rollout, see (\cite{abnar2022quantifying}). Briefly, attention rollout recursively computes the token attentions in each layer of a trained transformer model, resulting in an \textit{attention map}, analogous to class activation maps in CNNs.

\begin{table}[ht!]
\centering
\begin{threeparttable}
\caption{Hyperparameters used in the Vision Transformer parameter sweep.}
\label{table:VIT_hyperparameters}
\begin{tabular}{ll}
\toprule
\textbf{Hyperparameter}        & \textbf{Value}               \\ \hline
Batch size   & \textbf{16}, 32                            \\
Learning rate & 1e-5, \textbf{1e-4}, 1e-3                     \\
Depth & 4, 8, \textbf{16}\\
Number of Heads & 2, \textbf{4}, 8\\
Embedding Dimension & 128, \textbf{256}, 512\\
Patch size & \textbf{(7, 8)} \\
Kernel size & \textbf{(7, 8)}\\
Optimizer & \textbf{Adam} (\cite{kingma2014adam}) 
\\
Visualization Technique & \textbf{Attention-Rollout} (\cite{abnar2022quantifying}) \\
Loss Function & \textbf{Cross-entropy} \\ 
\bottomrule
\end{tabular}
\begin{tablenotes}
\small
\centering
\item \textbf{Bold values} indicate hyperparameters used in the final selected model.
\end{tablenotes}
\end{threeparttable}
\end{table}

\subsubsection{Support vector machine (SVM)}
SVM is a supervised learning model that classifies data by identifying the optimal hyperplane that separates data points into distinct classes. We implemented a linear SVM classifier that directly processes the registered fUSI data, without relying on predefined ROIs. In this setup, each pixel in the fUSI images is treated as an individual feature, allowing the SVM to perform unbiased feature selection across the entire imaging field. This approach enables the model to explore all aspects of the fUSI data, providing a comprehensive classification without any imposed spatial constraints.

\subsection{Training and Testing Data Sets}

All MK-801 vs. saline classification task models were subjected to 10-fold cross-validation for robust performance assessment. In each cross-validation fold, data from 5 out of 10 MK-801-injected animals and 7 out of 13 saline-injected animals was randomly selected for training, while the remaining data was held out for testing. This near 50-50 training/testing was chosen to ensure that a substantial number of animals were allocated to the test set for each cross-validation fold, which is particularly important for generating reliable class activation maps, feature weights visualizations, and attention maps. For each cross-validation fold, these visualizations are generated via averaging single-animal maps from the given testing set.
While increasing the training set size could improve classification accuracy, it would reduce the number of animals available for visualization, potentially introducing bias. Specifically, with fewer test subjects, the class activation maps/attention maps would be averaged over a smaller sample, making them less representative and more susceptible to individual variability. For smaller datasets, a larger proportion of the available data may need to be allocated for training to ensure the models learn effectively.

CNN and ViT models were trained on fUSI data from the final five minutes of post-injection recordings since our previous study shows that the effects of MK-801 on brain activity become more potent over time (\cite{crown2024theta}). Testing was conducted on the entire post-injection period using a one-minute sliding window. For SVM, we trained and tested a separate model on each successive one-minute interval to capture temporal variations in feature importance. This is done to compare the resulting \textit{feature weights visualization} with the class activation maps and attention maps from CNN and ViT, which are dynamic despite static training data. For a given SVM, the feature weights are determined solely by the data used to \textit{train} the model. To capture temporal variations in feature performance, we trained a new model for each one-minute interval. Each model generates a feature weight map corresponding to the SVM trained on that specific interval. These individual maps are then concatenated to create a comprehensive feature weights visualization, allowing for the analysis of how feature importance evolves over time.

\subsection{Model performance}

Model performance was assessed using 10-fold cross-validation. For each iteration, the model was trained on the designated training set and tested on the remaining test set, and a sliding window of one minute was utilized to analyze the model's performance over time. The adequacy of the model was evaluated using the following metrics:

\begin{itemize}
    \item Accuracy: (TP + TN)/ (P + N).
    \item Precision: TP / (TP + FP).
    \item Recall: TPR = TP/P.
    \item Area Under Curve (AUC): Area under the receiver operating characteristic (ROC), which is the plot of the true positive rate (TPR) against the false positive rate (FPR). FPR is defined as FP/N. 
    \item F1-Score: Harmonic mean of precision and recall: 2TP/(2TP + FP + FN).
\end{itemize}

Where P and N represent the number of positive and negative instances, respectively, and TP, TN, FP, and FN denote true positives, true negatives, false positives, and false negatives. To reduce potential biases from individual training/testing set selections, we averaged evaluation metrics in each temporal window across all cross-validation iterations. 

\subsection{Statistical Analysis}
For each model (CNN, SVM, and ViT), two-way repeated measure ANOVAs were conducted separately for MK-801-injected and saline-injected animals to evaluate differences in CBV changes between identified and non-identified regions during the post-injection period, with region (identified and non-identified) and time as factors, with data pooled across all animals within the MK/saline group.

A separate two-way repeated measures ANOVA compared classification performance over time among the three models during the same post-injection period, with model type (CNN, SVM, and ViT) and time as factors. Here, performance metrics for each model incorporated all cross-validation folds. For significant main effects of model type in the model comparison ANOVA, post-hoc pairwise comparisons were conducted using Tukey's test to determine which specific models differed significantly from each other, while controlling for the family-wise error rate in these multiple comparisons. All statistical analyses were conducted using Matlab R2020b.

\section{Results}

\subsection{Comparative model performance in detecting MK-801 effects}

\begin{figure}[h!t]
    \centering
    \includegraphics[width = \textwidth]{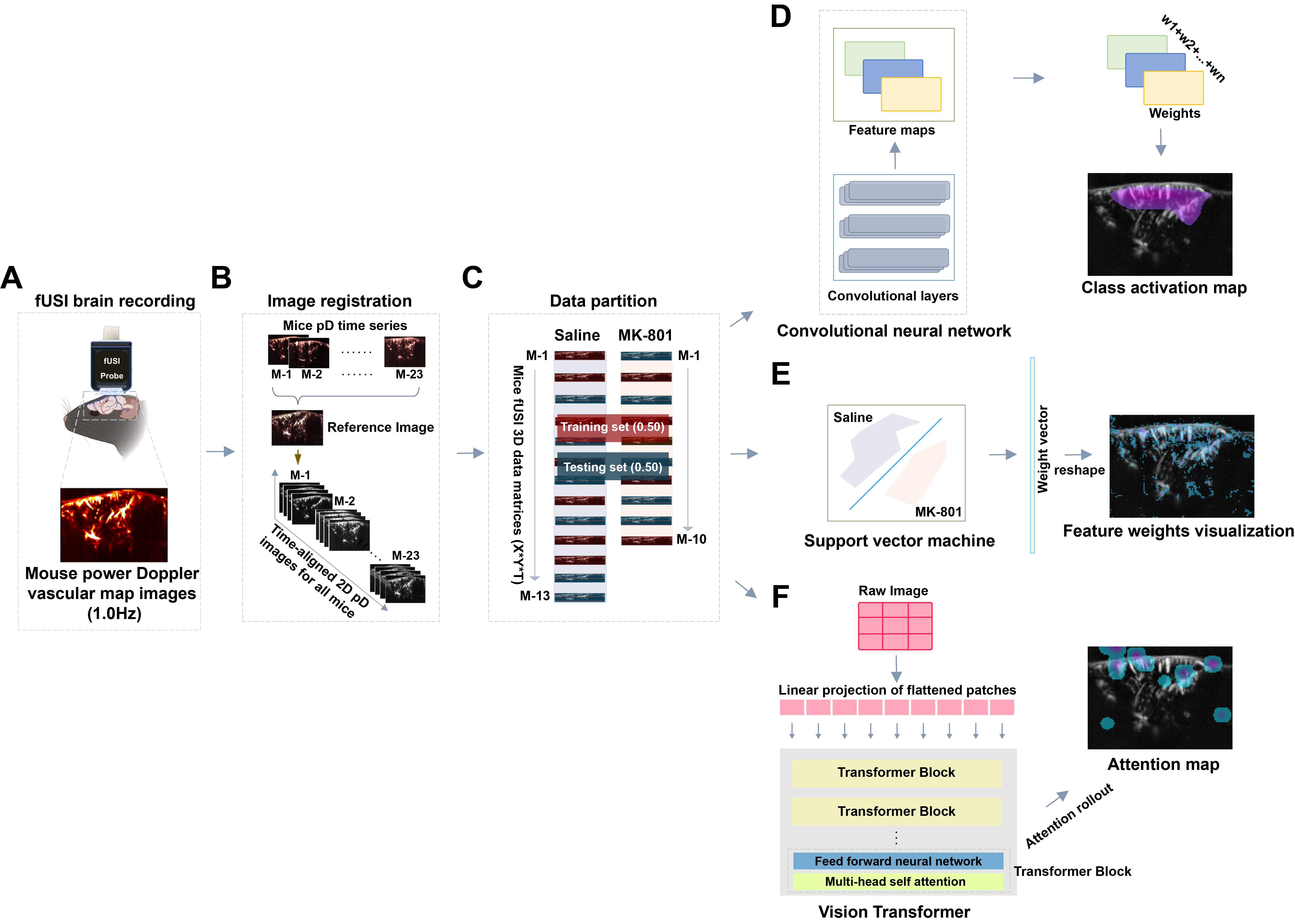}
    \caption{Overview of analysis combining fUSI and machine learning to detect MK-801 effects in the brain.
     (\textbf{A}) fUSI mice brain signal acquisition. (\textbf{B}) Mice 2D power Doppler (pD) image time series alignment to a common reference image. (\textbf{C}) Data partitioning. (\textbf{D}) CNN, feature selection and class activation mapping. (\textbf{E}) SVM and feature weights visualization. (\textbf{F}) ViT and attention map.}
     \label{fig:Flowchart}
\end{figure}

To investigate the spatiotemporal effects of MK-801 on brain hemodynamics, we developed an analytical framework that integrates fUSI with machine learning to classify drug-induced changes (Fig.~\ref{fig:Flowchart}). Power Doppler (pD) images were recorded from anesthetized mice receiving either MK-801 or saline injections. The analysis consisted of three main stages: data acquisition and preprocessing, model development and training, and performance evaluation (Fig.~\ref{fig:Flowchart}A-C). In the preprocessing stage, we implemented an advanced non-rigid image registration method {\it Imregdeform}. To establish a common reference frame, we selected a mouse with high-quality vascular maps to serve as the reference animal. We then created a stable reference image by averaging the first ten consecutive frames taken from the third minute of the five-minute baseline recording period from this animal. All images were subsequently aligned to this reference frame, ensuring the models did not rely on varying brain anatomical structures across animals for classification.

\begin{figure}[h!t]
    \centering
    \includegraphics[width=0.6\textwidth]{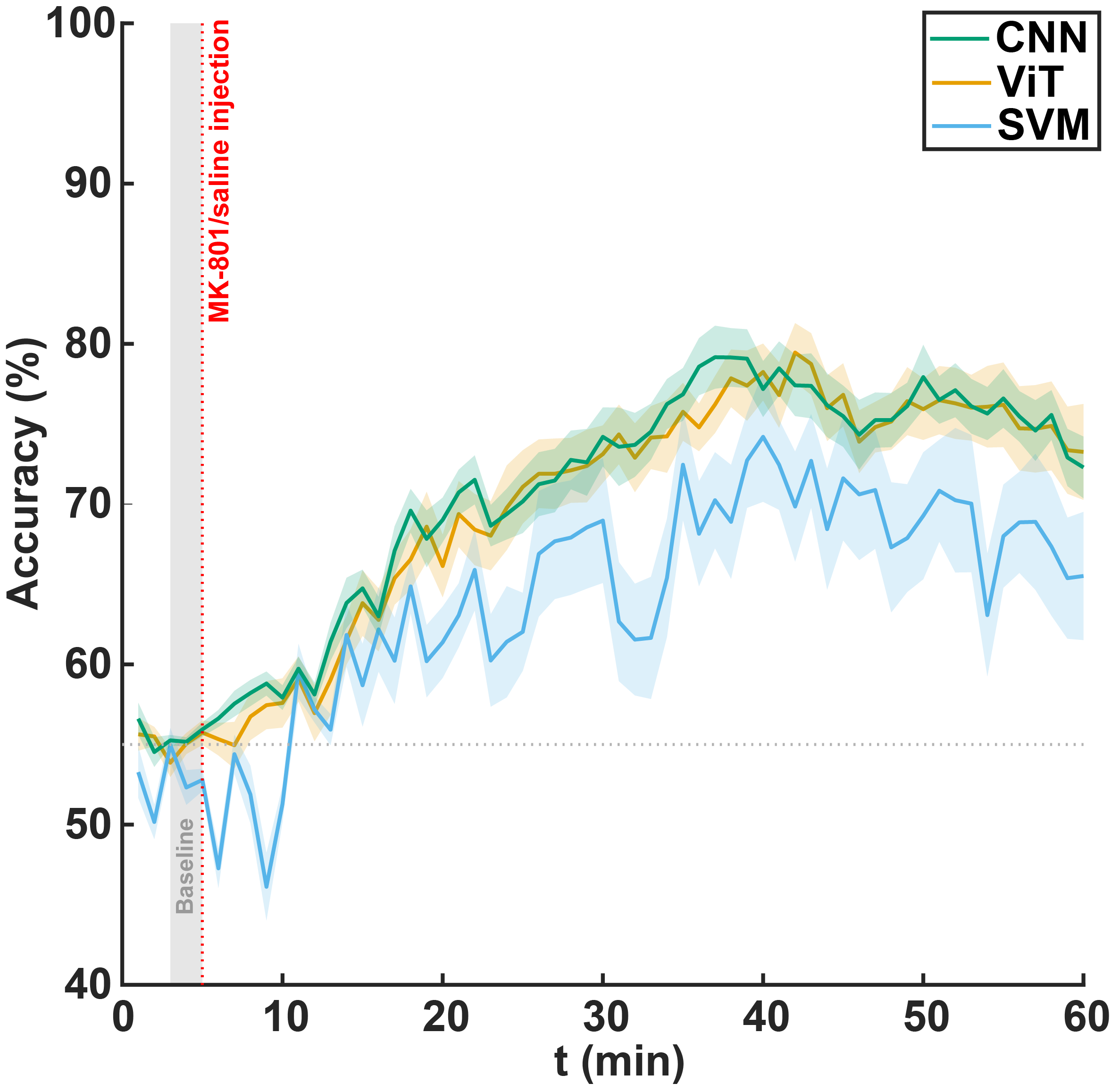}
    \caption{Model classification accuracy for distinguishing MK-801 group from saline group using ViT, SVM, and CNN. Shaded areas represent standard error. The vertical dotted line at 5 min marks the injection time, and the gray shaded area (minutes 3-5) indicates the baseline period.}
    \label{fig:accuracy}
\end{figure}

The CNN and ViT architectures underwent extensive parameter sweep through grid search followed by cross-validation of the top-performing configurations (Tables \ref{table:CNN_hyperparameters}, \ref{table:VIT_hyperparameters}), while we also included a linear SVM as a classical machine learning approach for comparison. We trained each model using pD images from the final five minutes of recording, a period in which MK-801 effects are expected to peak, then evaluated their performance across successive one-minute intervals spanning the entire 60-minute experiment using all cross validation folds. We assessed the ability of each model to distinguish the effect of MK-801 over time (Fig.~\ref{fig:accuracy}). All three models showed dynamic classification performance that evolved with the progression of MK-801 induced effects. Consistent with our hypothesis that classification accuracy would improve as the drug effects emerged, all models performed near chance level at the time of injection (CNN: 55.9 $\pm$ 1.4\%, ViT: 55.7 $\pm$ 2.4\%, SVM: 52.8 $\pm$ 2.2\%, mean $\pm$ SD across folds). As MK-801 gradually took effect, both CNN and ViT demonstrated robust performance improvements, with CNN reaching a peak accuracy of 79.2 $\pm$ 6.2\%  at 32 minutes post-injection, and ViT achieving a comparable peak of 79.5 $\pm$ 5.8 \%  at 37 minutes post-injection. SVM showed more modest performance, reaching a maximum accuracy of 74.2 $\pm$ 12.8 \%  at 35 minutes post-injection. Two-way repeated measures ANOVA performed on the post-injection period (minutes 5-60) revealed a significant main effect of method ($p < 0.01$) and time ($p < 0.001$) but no significant method $\times$ time interaction ($p = 0.96$). Post-hoc pairwise comparisons with the Tukey test showed that CNN significantly outperformed SVM (mean difference: 6.69\%, $p < 0.05$), while no significant differences were found between CNN and ViT (mean difference: 0.71\%, $p = 0.74$) or between SVM and ViT (mean difference: -5.98\%, $p = 0.13$). Additional model performance metrics including precision, recall, F1-score, and ROC-AUC are presented in Figure S1.

\subsection{CNN captures anatomically relevant MK-801 induced brain hemodynamic responses}

To ensure a fair comparison between models, we standardized the region selection approach by using a fixed area percentage threshold across all three methods. For each model's output (CAMs for CNN, feature weights for SVM, and attention maps for ViT), we first computed the mean across the duration of the experiment, and then selected the top 15\% of pixels using percentile-based thresholding. This approach addresses the inherent differences in value distributions across methods and provides an consistent basis for comparing their biological relevance. To validate the robustness of our findings, we performed sensitivity analyses across different threshold percentages (see videos 1-3 in Supplementary Materials), which showed that our conclusions remained consistent across a range of reasonable thresholds.

\begin{figure}[h!t]
    \centering
    \includegraphics[width=\textwidth]{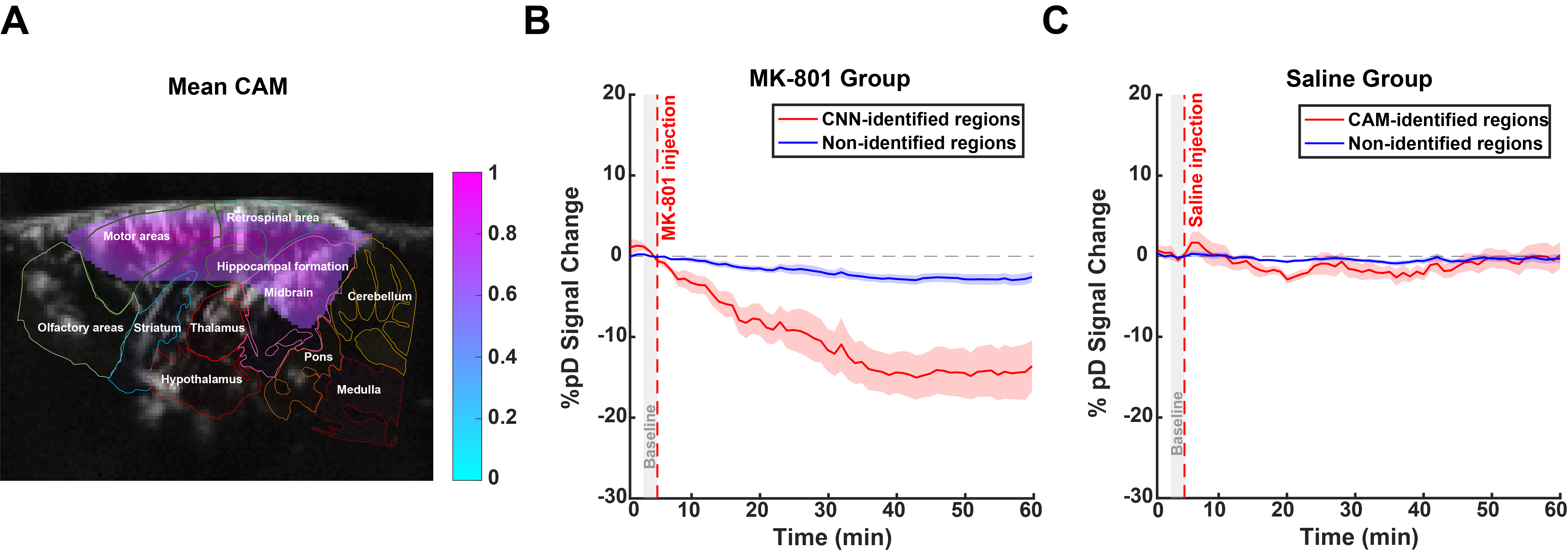}
    \caption{CNN-derived CAM and CBV dynamics in MK-801 and saline-injected animals. (\textbf{A}) Mean CAM for the MK-801 class overlay on the reference image, highlighting regions of importance in CNN-based classification of MK-801 effects. The top 15\% of pixels were selected using percentile-based thresholding. Brighter regions indicate higher relevance. (\textbf{B}) Time series of $\%pD$ (i.e., CBV) changes in CNN-identified regions (red curves) compared to non-identified regions (blue curves) for MK-801 group of animals.  (\textbf{C}) Similar to B but for the saline group of animals. The shaded areas represent standard error derived from averaging across animals. The red dashed line at 5 min marks the injection time. A gray patch (3-5 min) indicates the pre-injection baseline window.}
    \label{fig:CNN_analysis}
\end{figure}

The class activation maps for CNN (generated using XGRAD-CAM) revealed consistent activation patterns centered on cortical regions and extending into specific subcortical structures including the hippocampus and midbrain (Fig.~\ref{fig:CNN_analysis}A). These activation patterns offer an unbiased view on the brain regions that have a key role in  differentiating  the effects of MK-801 from saline control conditions. By focusing on the MK-801 group, the activation patterns reflect regions where the hemodynamic signal is strongly influenced by MK-801 administration. To validate the biological relevance of these CNN-identified regions, we measured CBV changes within these areas. The CBV changes were calculated as the percent change of pD signal relative to baseline - i.e., the mean pD signal during the last two minutes (minutes 3–5) prior to injection. This specific baseline window was chosen to ensure signal stability prior to drug administration. Notably, in MK-801-injected animals, CNN-identified brain regions exhibited a progressive decrease in CBV, reaching -13.63 $\pm$ 10.02\% (mean $\pm$ SD across cross animals) from baseline at 55 minutes post-injection, compared to -2.55 $\pm$ 2.04\% for non-identified regions - i.e., regions not identified by the CNN as affected by MK-801 administration. A repeated measures ANOVA performed on the post-injection period (5-60 minutes) revealed a significant main effect of region  ($p < 0.001$), a significant effect of time ($p < 0.001$) and a significant region $\times$ time interaction ($p < 0.001$), indicating that the CBV values significantly changed over the course of the experiment regardless of region type, and the temporal evolution of CBV changes was significantly different between regions (Fig.~\ref{fig:CNN_analysis}B). This interaction reflects the gradual divergence of CBV responses, with identified regions showing an increasing reduction over time. On the other hand, analysis of the same regions in saline-injected animals showed minimal changes at 55 minutes post-injection (0.20 $\pm$ 5.94\% in identified regions vs -0.19 $\pm$ 2.03\% in non-identified regions). A repeated measures ANOVA performed on the post-injection period (5-60 minutes) revealed no significant main effect of region ($p = 0.27$), and although there was a significant effect of time ($p < 0.001$), there was no significant region $\times$ time interaction ($p = 0.77$) (Fig.~\ref{fig:CNN_analysis}C). The spatial distribution of drug-sensitive regions aligns with previous findings demonstrating MK-801-induced CBV reduction in the hippocampus and medial prefrontal cortex (mPFC) (\cite{crown2024theta}), areas known for their high density of NMDA receptors (\cite{watanabe1994distinct}).

\subsection{SVM and ViT show limited spatial specificity}
\begin{figure}[h!t]
\centering
\includegraphics[width=\textwidth]{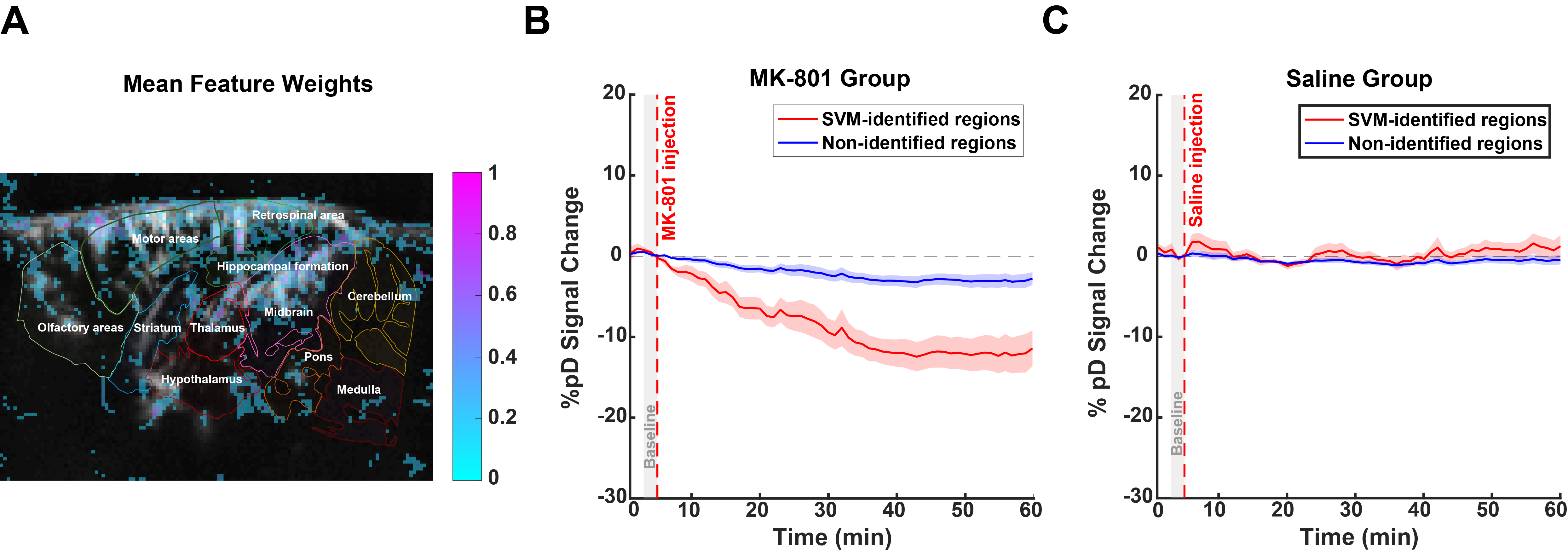}
\caption{SVM-derived feature weights and CBV dynamics in MK-801 and saline-injected animals. (\textbf{A}) Mean feature weights overlay on the reference image, highlighting regions of importance in SVM-based classification. The top 15\% of pixels were selected using percentile-based thresholding. Negative weights were extracted and inverted to specifically identify regions driving MK-801 (versus saline) classification, then normalized using percentile-based thresholding. Brighter regions indicate higher relevance for MK-801 detection. (\textbf{B}) Time series of $\%pD$ (i.e., CBV) changes in SVM-identified regions (red curves) compared to non-identified regions (blue curves) for MK-801 group of animals. (\textbf{C}) Similar to B but for the saline group of animals. The shaded areas represent standard error derived from averaging across animals. The red dashed line at 5 min marks the injection time. A gray patch (3-5 min) indicates the pre-injection baseline window.}
\label{fig:SVM_analysis}
\end{figure}

In contrast to CNN's anatomically specific activation patterns, SVM's feature weights revealed diffuse activation across the entire imaging field (Fig.~\ref{fig:SVM_analysis}A). In MK-801-injected animals, SVM-identified regions showed a significant decrease in CBV compared to non-identified regions, reaching -11.60 $\pm$ 7.04\% (mean $\pm$ SD across animals) at 55 minutes post-injection, compared to -2.91 $\pm$ 2.48\% in non-identified regions. A repeated measures ANOVA performed on the post-injection period (5-60 minutes) revealed a significant main effect of region ($p < 0.001$), a significant effect of time ($p < 0.001$) as well as a significant region $\times$ time interaction ($p < 0.001$), indicating that the CBV responses in SVM-identified regions differed significantly from non-identified regions throughout the recording (Fig.~\ref{fig:SVM_analysis}B). In contrast, while saline-injected animals showed significant differences between regions (main effect of region: $p < 0.05$) and changes over time (main effect of time: $p < 0.001$), they lacked the differential temporal evolution seen in the MK-801 group (no significant region $\times$ time interaction: $p = 0.99$). SVM-identified regions in saline-injected animals exhibited minimal CBV changes at 55 minutes post-injection ($1.30 \pm 4.94\%$) compared to non-identified regions ($-0.39 \pm 2.05\%$) (Fig.~\ref{fig:SVM_analysis}C). These findings indicate that while SVM successfully identified brain regions with distinct hemodynamic responses to MK-801, these regions were less spatially specific than those identified by CNN and CAM. SVM appears to capture meaningful temporal dynamics following MK-801 administration but relies on a broader, more diffuse spatial distribution of features, potentially reducing its anatomical specificity compared to the more focused CNN-based techniques.

\begin{figure}[h!t]
\centering
\includegraphics[width=\textwidth]{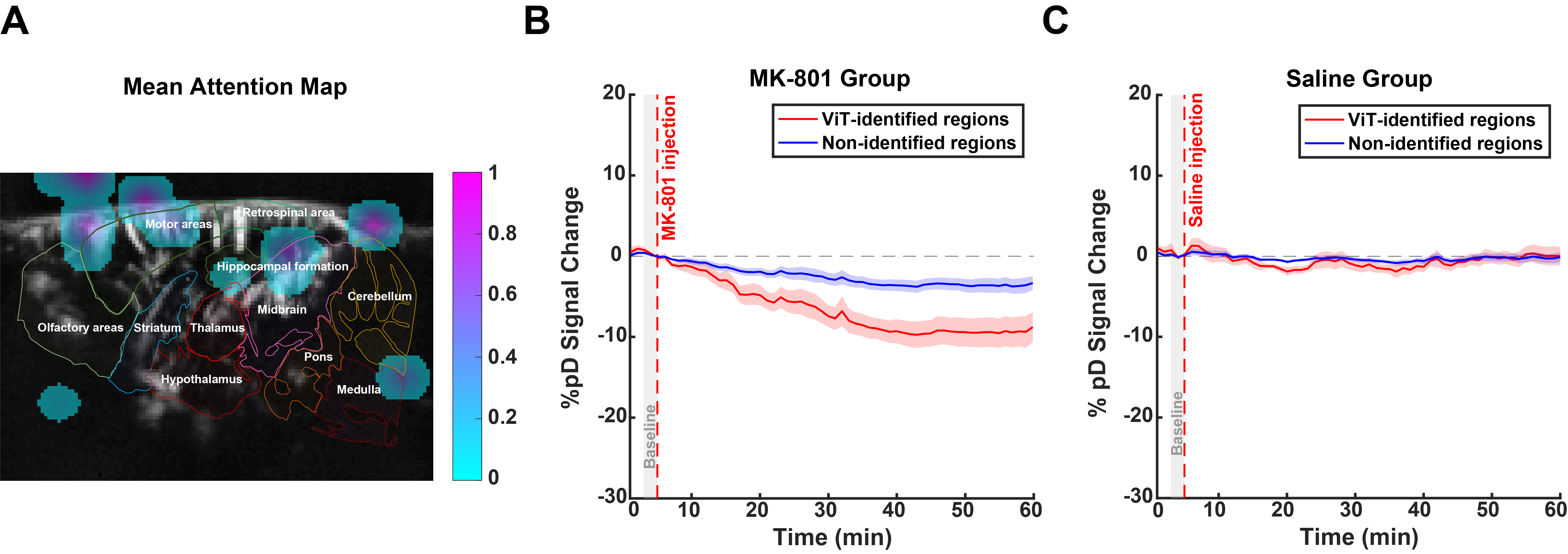}
\caption{ViT-derived attention maps and CBV dynamics in MK-801 and saline-injected animals. (\textbf{A}) Mean attention map for the MK-801 class overlay on the reference image, highlighting regions of importance in ViT-based classification. The top 15\% of pixels were selected using percentile-based thresholding. Brighter regions indicate higher relevance. (\textbf{B}) Time series of $\%pD$ (i.e., CBV) changes in ViT-identified regions (red curves) compared to non-identified regions (blue curves) for MK-801 group of animals. (\textbf{C}) Similar to B but for the saline group of animals. The shaded areas represent standard error derived from averaging across animals. The red dashed line at 5 min marks the injection time. A gray patch (3-5 min) indicates the pre-injection baseline window.}
\label{fig:ViT_analysis}
\end{figure}

ViT also detected drug-induced changes in brain hemodynamics, but its spatial activation patterns differed markedly from CNN and SVM (Fig.~\ref{fig:ViT_analysis}A). Focusing on the MK-801 attention map, in MK-801-injected animals, ViT-identified regions exhibited a substantial decrease in CBV, reaching -8.87 $\pm$ 5.86\% (mean $\pm$ SD across animals) at 55 minutes post-injection, compared to -3.39 $\pm$ 2.74\% in non-identified regions. A repeated measures ANOVA performed on the post-injection period (5-60 minutes) revealed significant effects of region ($p < 0.001$), time ($p < 0.001$), and a significant region $\times$ time interaction ($p < 0.001$), indicating that the CBV reduction was more pronounced in ViT-identified regions and evolved distinctly over time (Fig.~\ref{fig:ViT_analysis}B). In contrast, analysis of saline-injected animals showed no significant differences between ViT-identified regions (reaching -0.01 $\pm$ 4.10\% at 55 minutes post-injection) and non-identified regions (-0.16 $\pm$ 2.24\% at 55 minutes post-injection), with no significant main effect of region ($p = 0.36$) and no significant region $\times$ time interaction ($p = 0.99$), despite a significant effect of time ($p < 0.001$) (Fig.~\ref{fig:ViT_analysis}C). While ViT successfully identified regions showing differential drug responses, its attention maps included areas outside the brain structure (Fig.~\ref{fig:ViT_analysis}A), potentially limiting the biological interpretation of these findings. This ``checkerboard'' phenomenon has been highlighted as a drawback of attention rollout (\cite{achtibat2024attnlrp}). Together, these results suggest that while SVM and ViT could effectively distinguish between MK-801 and saline groups, they lacked anatomical precision in their feature selection compared to CNN's more anatomically precise identification of the corresponding brain regions.

% !TeX root = ../main.tex
\section{Discussion}

FUSI enables high-resolution measurement of CBV changes (\cite{mace2011functional,mace2013functional}). While its applications in neuropharmacology have grown (\cite{rabut2020pharmaco, Vidal2020pharmacofUS, Vidal2020drug,crown2024theta}), most analyses rely on predefined ROIs, which may miss important drug effects in other brain areas and introduce bias in result interpretation. To address these limitations, we developed and evaluated multiple machine learning approaches for automated, whole-brain analysis of drug-induced hemodynamic changes using MK-801 as a model NMDAR antagonist. 

Our systematic comparison of CNN, SVM, and ViT revealed distinct capabilities in analyzing fUSI data. CNN achieved both superior classification performance and biological specificity, demonstrating its unique ability to capture meaningful spatial patterns in brain hemodynamics. While SVM generated diffuse activation patterns and ViT exhibited attentions extending beyond brain boundaries, CNN demonstrated unique ability to capture meaningful spatial patterns in brain hemodynamics. The biological relevance of these CNN-identified regions is further validated by our recent comprehensive analysis using ROI-based methods (\cite{hakopian2025functional}), which independently showed that the hippocampus and mPFC exhibit the most pronounced CBV reductions and the greatest disruption in functional connectivity (hippocampus-mPFC pathway, t-score = -3.86). This convergence between our data-driven machine learning approach and ROI-based quantitative analysis provides strong validation that CNN can reliably identify biologically relevant brain regions without prior anatomical assumptions. Furthermore, the temporal progression of CNN performance, which peaks at approximately 32 minutes post-injection, corresponds with the period of pronounced connectivity disruption (25-30 min) transitioning to sparse network connectivity (35-40 min) (\cite{hakopian2025functional}), confirming the gradual, progressive nature of NMDAR antagonist effects on brain function.

\subsection{ROI-independent analysis}

The validation of CNN's biological specificity highlights the broader methodological advantages of our data-driven framework over traditional ROI-dependent approaches. Conventional analyses typically require manual or atlas-based ROI selection before signal extraction, which inherently constrains the analysis to predetermined regions. Our CNN-based technique evaluates spatial patterns across the complete imaging field, identifying relevant hemodynamic changes based on functional responses rather than anatomical assumptions. This methodology mitigates potential selection bias in neuropharmacological imaging analysis. The class activation mapping technique provides visualization of regions contributing to classification decisions, effectively generating data-driven ROIs based solely on functional responses to drug administration. This approach allows the detection of drug effects that might occur outside canonical regions associated with a drug's known mechanism of action.

\subsection{Limitations and future directions}

While our study demonstrates the potential of combining fUSI with interpretable machine learning techniques for investigating drug-induced changes in brain activity, it also opens avenues for technical refinement and further methodological exploration. The image registration performed using the {\it Imregdeform} algorithm provided sufficient alignment to detect drug-induced effects. Nonetheless, future optimization of registration parameters may further improve sensitivity to subtle hemodynamic change. Additionally, while our current analysis effectively captured drug-induced patterns, the interpretability of these models presents opportunities for advancement. For instance, future studies could consider ViT architectures with advanced interpretability techniques to overcome the ``checkerboard'' artifacts common in traditional attention visualizations as noted by (\cite{achtibat2024attnlrp}), potentially providing complementary insights while maintaining anatomical specificity.

The interpretation of our findings should take into account the anesthetic context under which they were obtained. Isoflurane use is an important factor to consider when comparing our findings with those from conscious animal studies, where MK-801 has been reported to increase CBV in rats (\cite{roussel1992acute}). In contrast, our observation of a CBV decrease aligns with results obtained under other anesthetics, such as halothane or $\alpha$-chloralose, where MK-801 has similarly been shown to reduce CBV (\cite{roussel1992acute, park1989effect}). These findings suggest that the cerebrovascular response to MK-801 is influenced by the anesthetic state. The known properties of isoflurane may add further complexity to this interaction: its vasodilatory effects (\cite{franceschini2010effect}) could alter baseline vascular tone, thereby modulating drug-induced hemodynamic changes, while its inhibitory action on NMDA receptors (\cite{nishikawa2000excitatory}) could interact directly with MK-801’s mechanism of action. By applying an identical anesthesia protocol to both control and treatment groups, our design allows for a valid comparison between conditions. Thus, the observed between-group differences likely reflect the impact of MK-801 within this anesthetized context. An important future direction will be to adapt this methodology for awake, freely moving animals to disentangle the intrinsic effects of MK-801 from those of anesthesia.

Looking ahead, the framework presented here holds considerable potential for broader application and eventual clinical translation. Its generalizability can be evaluated by extending it to other pharmacological agents and experimental models. Moreover, since fUSI has already been successfully applied to human brain imaging in various contexts (\cite{Imbault2017intra,demene2017functional,soloukey2020functional,agyeman2024functional,rabut2024functional,agyeman2025human}), our machine learning framework could be adapted to these clinical scenarios, providing novel insights into physiological processes, pathological conditions, and therapeutic interventions in the human central nervous system. While challenges exist—including the need for larger training datasets and potentially more sophisticated registration techniques—these represent technical rather than fundamental limitations. The minimally invasive nature, high spatiotemporal resolution, and cost-effectiveness of ultrasound make this technique favorable for clinical applications.

\subsection{Conclusion}
Overall, our study establishes a novel data analysis framework integrating deep learning with fUSI for neuropharmacological research. The proposed methodology offers an unbiased approach for detecting and localizing drug-induced changes without relying on predetermined ROIs. Our comparison of machine learning approaches demonstrates that CNNs provide superior biological specificity while maintaining classification performance, establishing a foundation for future fUSI data analysis. This methodological innovation has the potential to accelerate drug development by revealing previously overlooked drug effects and enabling more comprehensive characterization of therapeutic compounds across both preclinical and clinical settings.

\section*{Data and Code Availability}

The datasets generated and analyzed during the current study are available from the corresponding authors on reasonable request and after signing a formal data sharing agreement. The code is available on Github at  \url{https://github.com/DeightonJared/fUSI_CAM}.

\section*{Author Contributions}

\textbf{Jared Deighton}: Methodology, Software, Formal analysis, Visualization, Writing-Original Draft.
\textbf{Shan Zhong}: Methodology, Formal analysis, Visualization, Writing-Original Draft.
\textbf{Kofi Agyeman}: Methodology, Visualization, Writing-Original Draft.
\textbf{Wooseong Choi}: Investigation.
\textbf{Charles Liu}: Conceptualization, Writing-Review \& Editing, Funding acquisition.
\textbf{Darrin Lee}: Conceptualization, Writing-Review \& Editing, Funding acquisition.
\textbf{Vasileios Maroulas}: Conceptualization, Supervision, Writing-Review \& Editing, Funding acquisition.
\textbf{Vasileios Christopoulos}: Conceptualization, Supervision, Writing-Review \& Editing, Funding acquisition.

\section*{Acknowledgments}
This work has been partially supported by the Army Research Laboratory Cooperative Agreement No W911NF2120186, the Army Research Office W911NF-21- 1-0094, the Keck School of Medicine Dean’s Pilot Funding Program (DL- PI), the National Institute of Mental Health (NIMH: 1K08MH121757-01A1), the USC Neurorestoration Center, and the Marlan and Rosemary Bourns College of Engineering at the University of California Riverside through start-up funding. 

\section*{Declaration of Competing Interests} The authors declare no competing interests. 

\printbibliography

\end{document}